\documentclass{article}[10pt]

\RequirePackage{amsthm,amsmath,amsfonts,amssymb}
\def\bSig\mathbf{\Sigma}

\usepackage{booktabs}
\usepackage[margin=1in]{geometry}
\usepackage{setspace}
\usepackage{tabularx}
\usepackage{natbib}
\newcolumntype{Y}{>{\centering\arraybackslash}X}
\usepackage{graphicx}
\usepackage{url}

\title{Adjustment for Biased Sampling Using NHANES Derived Propensity Weights}

\author{Olivia M. Bernstein, University of California, Irvine\and 
Brian G. Vegetabile, RAND Corporation\and Christian R. Salazar, University of California, Irvine\and
Joshua D. Grill, University of California, Irvine\and
Daniel L. Gillen, University of California, Irvine}

\begin{document}

\maketitle





\begin{abstract}
The Consent-to-Contact (C2C) registry at the University of California, Irvine collects data from community participants to aid in the recruitment to clinical research studies. Self-selection into the C2C likely leads to bias due in part to enrollees having more years of education relative to the US general population. \cite{salazar_Racial_2020} recently used the C2C to examine associations of race/ethnicity with participant willingness to be contacted about research studies. To address questions about generalizability of estimated associations we estimate propensity for self-selection into the convenience sample weights using data from the National Health and Nutrition Examination Survey (NHANES). We create a combined dataset of C2C and NHANES subjects and compare different approaches (logistic regression, covariate balancing propensity score, entropy balancing, and random forest) for estimating the probability of membership in C2C relative to NHANES. We propose methods to estimate the variance of parameter estimates that account for uncertainty that arises from estimating propensity weights. Simulation studies explore the impact of propensity weight estimation on uncertainty. We demonstrate the approach by repeating the analysis by Salazar et al. with the deduced propensity weights for the C2C subjects and contrast the results of the two analyses. This method can be implemented using our \texttt{estweight} package in \texttt{R} available on GitHub.
\end{abstract}



\section{Introduction}
\label{s:intro}

Convenience samples are widely used to answer scientific questions because samples representative of the population may be impractical or unethical to collect. Most statistical methods assume representative sampling, but are naively applied to biased samples. Participant self-selection into convenience samples, may reduce the degree to which such samples are representative of target populations of interest leading to a biased sample. For example, the Consent-to-Contact (C2C) registry at the University of California, Irvine (\url{https://c2c.uci.edu}) enrolls potential participants to aid in clinical research recruitment strategies (\cite{grill_constructing_2018}). 
Participants are recruited into the registry through a variety of outreach strategies including emails, community talks, postcards, and other methods. 
Due to this the C2C is not expected to be representative of the United States (US) population. For example, the C2C participants tend to report more years of education relative to the general population, are more likely to be non-Hispanic White, have lower rates of comorbidities and higher rates of exercise (see Figure \ref{fig:covariates}). As part of the enrollment process the C2C participants self-report demographic and clinical characteristics. They also indicate whether they are willing to be contacted for potential participation in studies that involve various procedures or requirements such as lifestyle/behavioral modification, medication use, blood collection, brain imaging, autopsy, or lumbar punctures. Depending on the requirements and enrollment criteria of a study, participants who specifically report willingness to be contacted about required procedures are likely to be eligible and can be invited to participate, increasing the efficiency of recruitment. Due to the under-representation of racial/ethnic minority populations in clinical research (\cite{oh_diversity_2015}), members from our research team used data from the C2C to examine differences in willingness to participate by race/ethnicity (\cite{salazar_Racial_2020}). Self-selection into the C2C may, however, lead to biased estimates and limit generalizability of estimated associations. One way this bias could arise is if there were a differential relationship between race/ethnicity and willingness to participate by education level. 

Inverse probability weights are commonly used to adjust for design-based sampling of subpopulations and provide generalizable inference (\cite{lumley_Complex_2010}). For design-based survey sampling, inverse probability weights are generally prespecified and fixed by design as the inverse of the sampling probability for each unit. Weights can also be used to account for selection bias in convenience samples, but in this context they are not fixed or known. The population we want to obtain inference about is some target population. Convenience samples often do not reflect the desired target population because subjects self-select to participate. If the probability of selection into the sample was known, it would be possible to adjust for the selection mechanism and obtain generalizable inference. In this paper, we explore methods for estimating propensity to self-select into convenience sample weights using data from a national survey sample to better generalize inference from the C2C registry and propose a novel variance estimate for model parameters that accounts for uncertainty from propensity weight estimation.

Other methods incorporate outside information about a population to obtain more generalizable inference. MRP (multilevel regression and poststratification) derives subpopulation estimates from national surveys. 
Iterative proportional fitting (or raking) adjusts subpopulation counts to match known marginal counts. For examples of MRP see \cite{gelman_poststratification_1997} and \cite{park_bayesian_2004}, for an application of MRP see \cite{schell_state-level_2020} and for raking see \cite{bishop_discrete_1975} and \cite{little_Complete-case_2014}. 
In this paper, we focus on estimating inverse probability weights, or estimating propensity weights for a convenience sample. We propose to estimate propensity weights by combining a convenience and representative sample and estimating the probability of membership in the convenience sample versus the representative sample based on observed covariates. We further explore the impact of incorporating estimated propensity weights on bias, propose analytic variance estimates for weighted parameter estimates, and discuss the utility of using the National Health and Nutrition Examination Survey (NHANES) as a representative sample for estimating convenience sample membership probabilities for samples like the C2C. 

Our work is unique from previous research on this topic in several ways. (1) We investigate using NHANES as a representative sample to estimate convenience sample membership in biomedical applications. NHANES recruits approximately 5,000 individuals from across the US each year and over-samples people over 65 and minority groups, i.e. Hispanic, non-Hispanic (NH) Black and NH Asian subjects. Medical and dietary information are collected through structured questionnaires and in-person measurements (\cite{centers_for_disease_Control_and_prevention_Cdc_national_2013}). NHANES is a practical dataset for estimating propensity weights in biomedical convenience samples because it is representative of the US population, it contains medical measurements, and the data are open access (\url{https://wwwn.cdc.gov/nchs/nhanes/Default.aspx}). Most of the previous work on calibrating sampling weights assumed that a representative sample or a sample frame was readily available (as in \cite{zadrozny_learning_2004} and \cite{omuircheartaigh_generalizing_2014}). It is common to compare estimates of population parameters, such as the prevalence of diabetes in the US, to estimates from NHANES as a diagnostic check for selection bias (\cite{funk_electronic_2017}, \cite{greenblatt_factors_2019}, \cite{bailey_multi-institutional_2013}), but it is not often used for estimating sampling weights. 
(2) Concurrent work from \cite{ackerman_generalizing_2021} proposed the use of national survey samples, including NHANES, as a practical way to obtain a representative sample to generalize randomized trial results. They used sampling weights for the complex survey sample in the propensity weight estimation model and the final outcome model, but we utilize them as frequency weights and generate a representative pseudopopulation.
(3) We propose a novel analytic variance estimator that extends the sandwich estimator for survey weighted generalized linear models (\cite{lumley_fitting_2017}) to account for uncertainty from estimating the propensity weights. We follow a similar approach to \cite{schildcrout_longitudinal_2010} and treat the sampling weight estimation model and final outcome model as being simultaneously estimated. \cite{ackerman_generalizing_2021} instead proposed a double-bootstrap to account for uncertainty from estimating propensity weights and from the impact of non-response on the complex survey sample weights. The contribution of this paper is to provide a framework for utilizing NHANES to answer population questions via convenience samples and to provide further refinements for variance estimates of scientific outcome model parameters to obtain valid inference. 

The  manuscript is organized as follows: In Section \ref{s:methods}, we develop the proposed methodology for calibrating propensity weights and quantifying uncertainty in the final scientific model of interest. In Section \ref{s:simstudies}, we present a simulation study investigating the impact of estimated propensity weights on bias and variance. In Section \ref{s:application}, we apply our method to an analysis of racial and ethnic differences in research willingness. Finally, we conclude with a discussion of the advantages and limitations of the proposed method. 

\section{Methodology}
\label{s:methods}

Consider two collections of variables, $\mathbf{X_R}$ and $\mathbf{X_C}$ and let $\mathbf{X}$ be the set of variables in both $\mathbf{X_R}$ and $\mathbf{X_C}$, or $\mathbf{X}\equiv\mathbf{X_R}\cap\mathbf{X_C}$. Further, let $\mathbf{Y}$ be a subset of the variables in $\mathbf{X_C}$ that are not in $\mathbf{X_R}$, i.e., $\mathbf{Y}\in\mathbf{X_C}\setminus\mathbf{X}$. We will assume that it is possible to collect data sets from both $\mathbf{X_R}$ and $\mathbf{X_C}$, but that it is only possible to collect random samples from $\mathbf{X_R}$ which could be used to obtain population inference. A data set obtained from $\mathbf{X_C}$ will be ``convenience" and thus subject to potential bias (such as self-selection bias among other potential issues). Our ultimate goal is population-based inference on the set of variables in $\mathbf{Y}$. For example, we may be interested in the estimation of the association between some subset of variables, $\mathbf{Z}$, collected in $\mathbf{X}$, i.e., $\mathbf{Z} \subset \mathbf{X}$, and $\mathbf{Y}$. In our context $\mathbf{Z}$ is race/ethnicity (available in both NHANES and C2C) and $\mathbf{Y}$ is willingness to participate in research (available only in C2C). This is not possible using $\mathbf{X_C}$ alone and so our goal is to leverage data sets collected through random sampling ($\mathbf{X_R}$) and convenience ($\mathbf{X_C}$) to obtain valid population-based inference for $\mathbf{Y}$. 

To accomplish this goal, we construct weighted estimators for samples from $\mathbf{X_C}$ so that we can obtain population-based inference by constructing estimated weights, $w_C$, that leverage information about the differences in sample distributions between $\mathbf{X_C}$ and $\mathbf{X_R}$. Let $X$, $Y$, and $Z$ be the corresponding samples from the sets $\mathbf{X}$, $\mathbf{Y}$, and $\mathbf{Z}$. The general approach for estimating these weights is to collect data sets, $X_C$ and $X_R$ on $\mathbf{X_C}$ and $\mathbf{X_R}$, respectively. Our goal is to discriminate between the two datasets and to accomplish this we can concatenate $X_C$ and $X_R$ into a combined dataset. We then construct an auxiliary variable, $C_i$, which is an indicator for whether unit $i$ was collected in $X_C$. Thus, $C_i$ evaluates to 1 for units from $X_C$ and 0 otherwise. We can estimate the probability of a given unit arising from each sample and use this information to weight $X_C$ to be similar to $X_R$ and obtain inference on $\mathbf{Y}$. 

Let the subscript $i$ denote the observation for subject $i$. Our goal is to obtain a weighted estimator for population-based inference such that,
\begin{align}
\label{eqn:expY}
    E[w_{Ci}Y_i|C_i = 1] = E[Y_i].
\end{align}
Now, we define,
\begin{align}
\label{eqn:Pci}
    P_{Ci} = \textrm{Pr}(C_i = 1|X_i = x_i),
\end{align}
which allows us to construct weights
\begin{align}
\label{eqn:wci}
    w_{Ci} \propto \frac{1-P_{Ci}}{P_{Ci}}
\end{align}
so that Equation \ref{eqn:expY} holds under appropriate assumptions on the set $\mathbf{X}$. Equation \ref{eqn:Pci} is analogous to the propensity score from the causal inference literature (\cite{rosenbaum_central_1983}) and Equation \ref{eqn:wci} would correspond to the weights for the average treatment effect on the control (ATC) (\cite{li_balancing_2018}). Thus, we refer to the estimated weights $w_{C}$ as propensity for self-selection into the convenience sample weights, or propensity weights for short. 

\subsection{Assumptions}

 A propensity score is the probability of receiving treatment when conditioning on observed covariates (\cite{rosenbaum_central_1983}). There are three assumptions needed for estimating propensity scores and making causal conclusions: (1) unconfoundedness, (2) positivity, and (3) the stable unit treatment value assumption (SUTVA) (See Chapter 1 and 12 of \cite{imbens_Causal_2015} or Appendix 1 of \cite{greenland_Confounding_1999} for an overview). First, the unconfoundedness assumption requires that potential outcomes are conditionally independent of the treatment assignment given the observed covariates. Second, positivity requires that each unit has a positive probability of receiving both treatment and control. More formally, if $T$ is an indicator for receiving a treatment and $X$ are covariates, then $0 < \textrm{Pr}(T=1|X=x) < 1$ for all subjects. Third, SUTVA requires that each subjects treatment assignment does not affect any other subject's potential outcomes and there is no hidden variability in the treatment. 
 
 We make analogous assumptions in our context with propensity weights. The three analogous assumptions are as follows. First, we assume unconfoundedness, that the response is independent of the selection probability conditional on the collected covariates. In other words, any covariate (or a proxy of the covariate) related to selection must be measured in both the representative and biased samples. We are unable to balance on unmeasured covariates. Second, each subject must have a non-zero probability of being selected into the convenience sample, or $0 < \textrm{Pr}(C=1|X=x) < 1$. Although C2C registration is open to any adult, participants are primarily enrolled from Southern California. Although theoretically possible, the probability of people from outside of the Southern California region being sampled in the C2C is close to zero. Thus, we must assume that the relationship between race/ethnicity and research willingness does not vary by state within the US. SUTVA requires that someone else participating in either the C2C or NHANES does not affect any other subject's willingness to be contacted about research studies. This was unlikely to occur for subjects whom did not know each other, although NHANES does sample multiple units within a household. In the C2C, recruitment of participants from community events and other outreach events are similarly done in groups. Early C2C participants were encouraged to recruit their friends, but this practice was not very effective and thus should not result in a large violation of SUTVA. In practice, it is important to carefully design the data collection to include any covariates hypothesized to be related to the sampling probability. If there are any missing covariates, accounting for the sampling bias due to measured covariates should be better than ignoring the selection mechanism completely (\cite{masten_identification_2018}), but the unconfoundedness assumption is not testable (\cite{imbens_Causal_2015}). The non-zero sampling probability assumption should motivate researchers to collect participants from each subpopulation based on variables related to selection. We can upweight underrepresented subpopulations, but we are never able to learn about subpopulations that were never studied.


\subsection{Propensity Weight Calibration}
\label{s:htcalibration}

We start by estimating propensity weights for the convenience sample $X_C$ with $n_C$ observations. Let $m$ and $p$ be the number of variables in $\mathbf{X}$ and $\mathbf{Z}$, respectively. We include a column of 1s in $\mathbf{X}$ and $\mathbf{Z}$ to be able to estimate an intercept. When using a complex survey sample, such as NHANES, as the representative sample we need to first incorporate design weights to ensure it is representative of the population of interest because certain subpopulations may be oversampled by design. Let $X_S$ be a survey sample with $n_S$ observations and $\pi_{Si}$ denote the sampling probability for subject $i$ in the survey sample. In NHANES, the sampling probabilities for each subject account for both the survey design and both item and subject level non-response. To obtain a representative dataset, we utilize frequency weights, $w_{si} = \pi_{si}^{-1}$, which represent the number of subjects each sampled subject represents in the population and replicate each subject according to their frequency weight. We divide each frequency weight by the smallest observed weight and take the ceiling of it to obtain the number of replications: $w_{Si}^* = \textrm{Ceiling}(w_{Si}/\textrm{min}[w_S])$. We implement the frequency weights to obtain a representative sample $X_{R}$ with dimension $n_{R}\times m$, where each subject from $X_S$ is replicated $w_{Si}^*$ times for a total of $n_{R} = \sum_{i=1}^{n_S} w_{Si}^*$ observations. 

Recall, that $X$ is the sample of variables collected in both $X_r$ and $X_c$ and containing observations from both representative and convenience sample subjects. $X$ is an $n\times m $ dimension matrix where $n = n_R + n_C$. For notational convenience, let $\mathcal{C}$ be the set of subjects from the convenience sample with $|\mathcal{C}| = n_C$ and and $\mathcal{R}$ be the set of subjects from the representative sample with $|\mathcal{R}| = n_R$. To obtain $X$ in practice, we concatenate the convenience and representative samples for the variables in $\mathbf{X}$. We can derive the indicator for membership in the convenience sample, $C$, and append it to $X$. To estimate the propensity weights, $w_C$, defined in Equation \ref{eqn:wci} we can directly estimate the probability of convenience sample membership, $P_{Ci}$, defined in Equation \ref{eqn:Pci}. Many types of propensity weight estimation methods can be considered and we  discuss their advantages and disadvantages in the following section. 

\subsection{Classes of Propensity Weight Estimation Methods}
\label{s:weightmethods}

In this section we describe different sampling weight estimation strategies and provide examples of each that we will use as test cases. We discuss likelihood based methods and use logistic regression as an example, covariate balancing methods with the covariate balancing propensity score and entropy balancing as examples, and algorithmic methods with random forest as an example. We will explore these four examples of methods for estimating propensity weights and their impact on covariate balance of convenience samples and on bias and variance of estimated associations. 

Likelihood-based methods such as linear regression, logistic regression, probit regression, and penalized regression minimize the negative log-likelihood. We focus on logistic regression which takes the form $\mbox{logit}(P_{Ci}) = {X}_i\gamma$,  where $\mbox{logit}(\cdot)$ is the logit function, $\gamma$ is a $m \times 1$ vector of regression coefficients and ${X}_i$ is the $m \times 1$ vector of observed covariates for subject $i$. When implementing logistic regression, we include second order terms and use forward-selection with Akaike's Information Criterion (AIC) for selecting predictors using the \texttt{step} function in the \texttt{stats} package.

By definition, propensity scores are balancing scores (\cite{rosenbaum_central_1983}) and another strategy for propensity weight estimation is to directly balance covariate distributions across the two classes of a binary outcome. Covariate balancing methods, such as the covariate balancing propensity score (CBPS) and entropy balancing (EB), optimize weights by directly targeting covariate balance between the the convenience sample ($C_i = 1$) and the representative sample ($C_i = 0$). 

CBPS extends the logistic regression model by incorporating additional moment balancing constraints (\cite{imai_Covariate_2014}). In our context, estimated propensity weights should balance the covariate distribution of the convenience sample so it matches the representative sample (\cite{li_balancing_2018}). Thus, we  estimate weights for the average treatment effect on the control (ATC) so that the representative sample is the reference population. In addition to solving the estimating equations, the CBPS includes covariate balancing conditions, that is, 
\begin{equation}
\label{eqn:balancing}
    E \Bigg\{ \frac{(1-P_{Ci})C_{i}f(X_{ij})}{P_{Ci}} - {(1-C_{i})f(X_{ij})}\Bigg|X_{ij} = x_{ij}\Bigg\} = 0,
\end{equation}
where $f(X_{ij})$ is a function of $X_{ij}$. For example, $f(X_{ij}) = (X_{ij}; X_{ij}^{2})$  would ensure the first and second moments of each covariate will be balanced. We fit the CBPS model with the \texttt{CBPS} package and included balancing constraints on second order orthogonal polynomial terms. 

Entropy balancing is a non-parametric approach to weight estimation that incorporates moment balancing conditions into model selection (\cite{hainmueller_entropy_2012}). Entropy balancing allows for the inclusion of initial base weights that contain information about population prevalence. Entropy balancing estimates weights that minimize the divergence between estimated weights and base weights. If there are no base weights available, they can be treated as uniform for all units ($b_{i} = 1/n)$. Similar to the CBPS, we want to estimate weights for the ATC where the representative sample is considered the reference group. Assuming Kullback–Leibler divergence, EB minimizes
$H(w_{Ci}) = \sum_{i \in \mathcal{C}} w_{Ci} \log(w_{Ci}/b_{i})$,
subject to balancing constraints,
\begin{equation*}
    \frac{1}{n_R}\sum_{i\in\mathcal{R}}x_{ij}^d = \sum_{i\in\mathcal{C}}w_{Ci}x_{ij}^d \textrm{, for }d = 1,...,D,
\end{equation*}
where $x_{ij}$ is the $j$-th covariate for subject $i$. 
Additionally, the optimization is subject to normalizing constraints (that can be relaxed) ensuring that the estimated weights must sum to one and be non-negative. We implemented entropy balancing with $D = 3$ using the \texttt{entbal} package for \texttt{R} (\url{https://github.com/bvegetabile/entbal}) available on GitHub (\cite{vegetabile_nonparametric_2021}).

Unconstrained algorithmic methods, such as support-vector machines or random forest, do not require specifying a model. They predict class probabilities that minimize the out-of-sample prediction error and do not assume a distribution or any moments of the response. As an example, we focus on random forest (\cite{breiman_Random_2001}) which is a flexible model that does not require the user to specify a functional form of predictors. RF is an extension of classification and regression trees or CART (\cite{breiman_Classification_1984}) and limits susceptibility to overfitting by introducing stochasticity. RF builds separate decision trees on bootstrapped samples and averages the prediction across trees for each subject, a technique known as bagging, and for each node of a decision tree, only a random sample of predictors are considered for splitting. To prevent extreme weight values, we trim RF estimates of the probability of convenience sample membership that are 0 or 1 and replace them estimate with 0.01 and 0.99, respectively (\cite{lee_weight_2011}). We fit RF using the \texttt{randomForest} package in \texttt{R} (\cite{liaw_Classification_2002}).

To demonstrate the advantages and disadvantages of the different approaches, we will evaluate the performance of the different propensity weight estimation methods described above: logistic regression, covariate balancing propensity score (CBPS), entropy balancing (EB), and random forest (RF). 

 \subsection{Quantifying Uncertainty}
\label{s:methodsvar}

There are several common ways to estimate the variance of coefficients from models with weights. Standard analytic variance estimates for coefficient estimates from weighted GLMs are an extension of the sandwich, or Huber-White, estimator (\cite{freedman_so-called_2006}). These design based variance estimates are included in most survey sampling software packages such as the \texttt{svyglm} function in the \texttt{survey} package  (\cite{lumley_fitting_2017}). This approach assumes the propensity weights are fixed, i.e. not estimated, but uncertainty from propensity weight estimation will likely impact the uncertainty of the parameter estimates in the scientific model. If there is little uncertainty in $\widehat{w}_{C}$, then the design based errors used in survey sampling methodology will likely perform well. Alternatively, resampling methods, such as the bootstrap, are more computationally intensive, but can be used to account for the  impact of the uncertainty in the propensity weight estimation procedure on the variance of the parameter estimates by reestimating weights within each bootstrap sample. 

In this section, we derive an analytic variance estimate that accounts for uncertainty from the propensity weight estimation method. We use a simultaneous estimating equation approach for variance estimation and extend the approach of \cite{schildcrout_longitudinal_2010}. We  treat both the propensity weight estimation and final scientific model as if they are being estimated simultaneously and derive the sandwich estimator for the parameters of the scientific model. Of the four propensity weight estimation methods we have considered, only the logistic model allows for a readily tractable analytic variance estimate that accounts for uncertainty in propensity weights. The design based variance estimate can be used for other weight estimation methods.

Suppose the scientific outcome model with response, $Y_i$, and the $p\times 1$ vector of covariates, $Z_i$, for subject $i$ is $\eta(\mu_i) = z_i \beta$
where $\eta(\cdot)$ is a link function, $\mu_i = E(Y_i|Z_i=z_i)$, and $\beta$ is a $p \times 1$ vector of parameters. Recall, when applying this method to the analysis of the C2C, $Z$ is the subset of covariates in the C2C sample needed for the final analysis, which includes race/ethnicity and adjustment variables, and $Y$ is the willingness to participate. Assuming a representative sampling scheme, the $i$-th observation's contributions to the $k$-th element of the score equation is given by
\begin{equation*}
    U_{ki}(\beta) =  \Big( \frac{\partial \mu_i}{\partial \beta_k} \Big) \Big( \frac{Y_i-\mu_i}{V(\mu_i)} \Big ) = 0 \\
\end{equation*}
where $V(\mu_i) = V(Y_i)$ and for $k = 1...p$ (\cite{nelder_generalized_1972}). When using an unrepresentative sample, each subjects contribution to the score is weighted by their propensity weight, $w_{Ci}$ as follows,
\begin{equation*}
     \bar{U}_{k}(\beta) =  \sum_{i\in\mathcal{C}} w_{Ci} U_{ki}(\beta).
\end{equation*}
Define the logistic regression propensity weight model, $\psi_i$, as $\psi_i = \mbox{logit}(P_{Ci}) = {x}_i \gamma$ where $\gamma$ is a $m \times 1$ vector of coefficients. Let $x_i$ be the $i$-th row of $X$, the combined $n \times m$ matrix including covariates from the convenience sample and representative sample. The estimated propensity weights $w_{Ci}$ are a function of the probability of convenience sample membership (Equation 1). The $m \times 1$ score equation has element $j$,
\begin{align}
\label{eqn:logisticscore}
    T_j(\gamma) = \sum_{i\in \mathcal{C}\cup\mathcal{R}}T_{ji}(\gamma) = \sum_{i\in \mathcal{C}\cup\mathcal{R}} (C_{i} - P_{Ci})x_{ij} = 0 ,
\end{align}
where $j = 1...m$. We consider both the score equation for the propensity weight estimation, $T_i(\gamma)$, and the score equation for the scientific outcome model, $\bar{U}_{i}(\beta,\gamma) = \bar{U}_{i}(\beta)$. We include $\gamma$ in the notation to emphasize that the score for the final scientific outcome is a function of the $\gamma$ through the propensity weights. To simplify notation, we sometimes refer to $T_i(\gamma)$ and $\bar{U}_{i}(\beta,\gamma)$ as $T_i$ and $\bar{U}_{i}$, respectively. We combine the two estimation equations into a stacked estimating equation 
\begin{align}
\label{eqn:stacked}
\begin{pmatrix}
\sum_{i\in \mathcal{C}\cup\mathcal{R}}T_i(\gamma) \\
\sum_{i\in \mathcal{C}}\bar{U}_{i}(\beta,\gamma) 
\end{pmatrix}
=0.
\end{align}
Using a first order Taylor series expansion of the stacked estimating equation (Equation \ref{eqn:stacked}) we obtain the variance estimate,
\begin{align}
\label{eqn:Vprop}
    \widehat{{V}}_{Prop}[(\widehat\gamma, \widehat\beta)] = \widehat{I}^{-1}\widehat{Q}\widehat{I}^{-1},
\end{align}
where the parameters have been replaced with maximum likelihood estimates. In Equation \ref{eqn:Vprop}, $I$ is Fisher's information matrix under the assumed distributions of $Y_i$ and $C_i$ such that,
\begin{align*}
    I = 
    \begin{pmatrix}
    I_{TT} & 0 \\
    I_{UT} & I_{UU}
    \end{pmatrix}
\end{align*}
and $Q$ is the true variance of the score, where
\begin{align*}
    Q = \textrm{Var}
    \Bigg(
    \begin{matrix}
\sum_{i\in \mathcal{C}\cup\mathcal{R}}T_i(\gamma) \\
\sum_{i\in \mathcal{C}}\bar{U}_{i}(\beta,\gamma) 
\end{matrix} \Bigg| X = x, Z=z \Bigg) =
    \begin{pmatrix}
    E[TT^T|X=x] & R^T \\
    R & E[\bar{U}\bar{U}^T|Z=z]
    \end{pmatrix}.
\end{align*}
Derivations of the components of $I$ and $Q$ are provided in Appendix \ref{s:mathderiv}. The reader may notice similarities to the design based variance used in the survey sampling literature without the finite population correction factor (\cite{lumley_fitting_2017}). Define $\widehat{A} = \widehat{I}_{UU}$ and $\widehat{B} = \bar U \bar U^T$ so the proposed variance estimator is, 
\begin{align*}
   \widehat{V}_{Prop}(\widehat\beta) &=  \widehat{A}^{-1}\widehat{B}\widehat{A}^{-1} - \widehat{A}^{-1}\widehat{I}_{UT}\widehat{I}_{TT}^{-1}\widehat{R}^T \widehat{A}^{-1}.
\end{align*}
Thus the proposed variance estimate can be expressed as the standard design based variance estimator plus a correction factor. We have provided a \texttt{estweight} package available for \texttt{R} on GitHub (\url{https://github.com/oliviabern/estweight}). The \texttt{estweight} function takes a representative sample, convenience sample, and the final outcome model and provides weighted parameter estimates. If the user selects a logistic propensity weight estimation method, the function returns the proposed variance estimate, otherwise it provides standard design-based variance estimates.

\section{Simulation Studies}
\label{s:simstudies}

We considered the impact of our proposed weight estimation on bias of estimated associations and the accuracy of uncertainty quantification procedures through empirical simulation studies. We designed a simulation study to be similar to the analysis of \cite{salazar_Racial_2020}. NHANES collects cross-sectional data on a 2-year cycle so we combined data from the 2013-2014 and 2015-2016 surveys. All simulations utilized the NHANES data, and like Salazar et al., we excluded all subjects with a reported race or ethnicity of ``other" for a total of $n_S = 4,471$ subjects. All subjects had complete data on age, sex, education, race, ethnicity, medical history (high blood pressure, diabetes, kidney disease, liver disease, coronary heart disease, cancer, major depression, prescription drug use), exercise, and amount of sleep. To obtain a representative sample we replicated each observation according to their frequency weight for a final sample size of $n_{R}=38,811$ observations.  We refer to this representative dataset as NHANES-REP. Code for creating NHANES-REP and for reproducing the simulation study is available on GitHub (\url{https://github.com/oliviabern/estweight\_simulationstudy}).

\subsection{Simulation Set Up}
\label{s:simsetup}

To investigate the potential impact of underrepresentation in samples and estimated propensity weights on bias and variance of estimated associations we used NHANES-REP as a finite population and drew both representative and deliberately biased samples. Subjects who are Hispanic, NH Black, NH Asian, or who have lower education levels and do not exercise tend to be underrepresented in the C2C and so we generated smaller sampling probabilities for these subpopulations. In this section we use $I$ to denote an indicator variable and this should be differentiated from our earlier use of $I$ as an information matrix. Let $P_{Ci}$ be the biased sampling probability for subject $i$ where $\mbox{logit}(P_{Ci}) = \psi_i$ with
\begin{align*}
    \psi_i &= .15I_{Female,i} + .25I_{High School,i} + .1I_{<High School,i} + .4I_{Some College,i} + .85I_{Hispanic,i} + .45I_{NH Asian,i} \\
    &+ I_{NH Asian,i}I_{Some College,i} +  .05I_{NH Black,i} + .75I_{NH Black,i}I_{Exercise,i} 
            -.001Age_i^2 + 4.
\end{align*}

Within each simulation, we drew a representative simple random sample of size 500 and a biased sample of size 500 with sampling probabilities $P_{Ci}$. We simulated $Y_i\sim \textrm{Bernoulli}(\pi_i)$ with $\mbox{logit}(\pi_i) = \rho_i$ and 
\begin{align*}
    \rho_i &= 1 + log(2)I_{Hispanic,i}  -log(3)I_{NH Asian,i} + log(1.5)I_{NH Black,i} -log(2)P_{Ci} \\
    &+ log(2)I_{Hispanic,i}P_{Ci} + log(4)I_{NH Asian,i}P_{Ci}  -log(3)I_{NH Black,i}P_{Ci}.
\end{align*}
We estimated propensity weights for subjects in the biased sample with each of the four propensity weight estimation methods described in Section \ref{s:weightmethods}. Similar to our applied example, we were interested in a model of the the marginal relationship between race/ethnicity where,
\begin{equation*}
    \mbox{logit}(\textrm{Pr}[Y_i = 1|X_i=x_i]) = \beta_0 + \beta_1 I_{Hispanic,i} + \beta_2 I_{NH Asian,i} + \beta_3 I_{NH Black,i}.
\end{equation*}
For each simulation, we (1) fit the above model in the representative sample with the objective of obtaining a similar estimate using a biased sample. (2) We then fit the model in the biased sample without any weighting, (3) with the true propensity weights ($w_{Ci} \propto \pi_i^{-1}(1-\pi_i)$), and (4) with the estimated propensity weights from each of the four estimation methods and compared the estimates to those obtained in the representative sample. We computed and compared analytic and bootstrap estimates of the standard error to the empirical Monte Carlo standard error. To prevent under-representation of small subpopulations in bootstrap samples, we used race/ethnicity as a stratification variable for sampling. When stratifying the bootstrap sample failed to provide adequate representation of subpopulations leading to extreme weights and inestimable coefficients, we removed the parameter estimates from bootstrap variance estimates. We conducted 1,000 simulations and used 200 bootstrap samples within each simulation.

\subsection{Simulation Results}

\subsubsection{Coefficient Estimates}
\begin{figure}
    \centering
    \includegraphics[width=\textwidth]{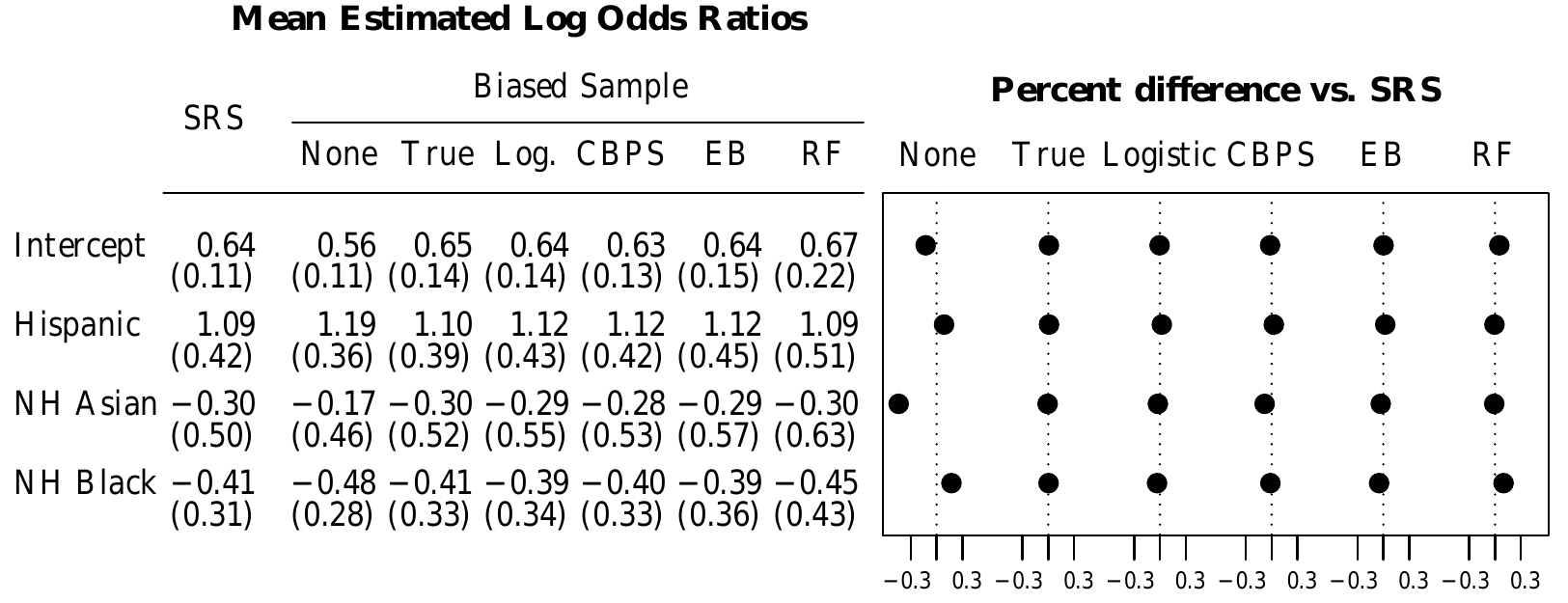}
    \caption{Results for the simulation study described in Section \ref{s:simsetup}. The average estimated log odds ratios (empirical standard errors) are presented in the table on the left. Estimates for the marginalized model fit in the simple representative sample (SRS) are in column 1 and estimates fit in a biased sample along with different types of propensity weights are in the other columns. Results compare models fit in a biased sample that do not include weights (None), incorporate the true propensity weights (True), or incorporate propensity weights estimated with a logistic (Log.), covariate balancing propensity score (CBPS), entropy balancing (EB), or random forest (RF) approach. Percent bias comparing average estimates fit in a biased sample to estimates fit in a simple random sample (SRS) are presented in the figure on the right.}
    \label{fig:orest}
\end{figure}

The average log odds ratios for representative and biased samples for each of the propensity weight types are shown in Figure \ref{fig:orest}. The goal of incorporating estimated weights is to match estimates fit using a biased sample (columns 2-7 of the tabulated results) to those estimated using a representative sample (column 1). Note that estimates derived from a biased sample that fail to account for the sampling scheme (i.e. no weighting, column 2) did not match those from the representative sample. Incorporating both true (column 3) and estimated (columns 4-7) propensity weights allowed us to match the representative sample estimates. The type of propensity weight model did not  have an appreciable impact as all of the weighted estimates did not vary much in comparison to the representative sample and to each other. Logistic regression performed well and allowed us to obtain weighted estimates fit in a biased samples that matched the association in the target population. It is a practical choice because it is parsimonious and easy to implement. It is important to note that the true sampling probabilities were generated from a logistic model, but the model included interactions that were not in the scope of the logistic model we used for estimation. In this example, incorporating estimated propensity weights was an effective method for obtaining inference on the target population.  

\subsubsection{Uncertainty Estimates}

\begin{figure}
    \centering
    \includegraphics[width=\textwidth]{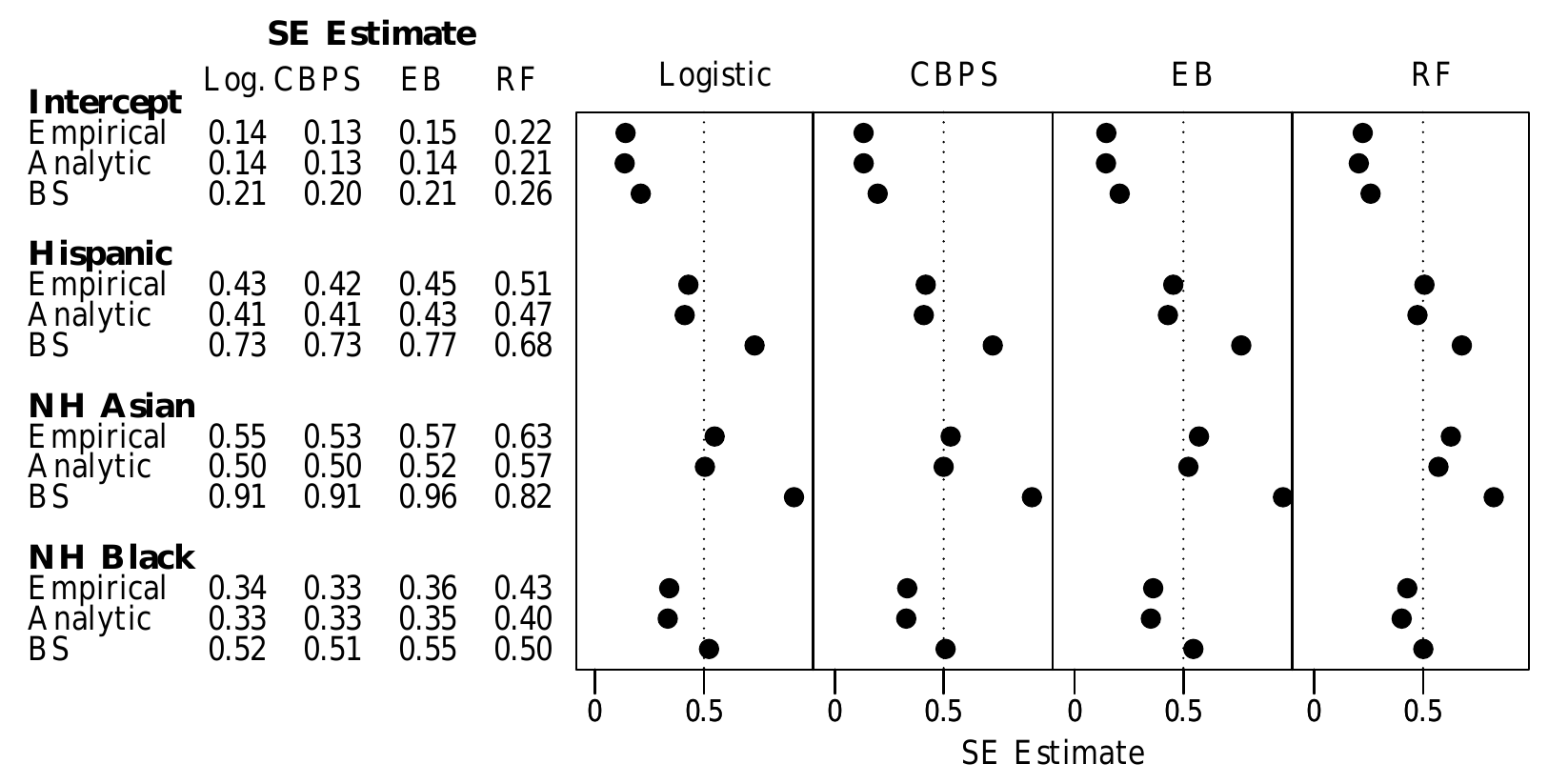}
    \caption{Standard error estimates for the simulation study described in Section \ref{s:simsetup}. The different standard error estimates (empirical, analytic and bootstrap) are reported for the coefficient estimates in the marginalized model fit in a biased sample. Standard error estimates are reported for models implementing propensity weights estimated using a logistic (Log.), covariate balancing propensity score (CBPS), entropy balancing (EB) or random forest (RF) method. Details of the standard error calculations are reported in Section \ref{s:methodsvar}}
    \label{fig:sesim}
\end{figure}

In this simulation, empirical standard errors were larger for weighted estimates fit in biased samples compared to unweighted estimated fit in a representative sample (Figure \ref{fig:orest}). We also investigated the impact of the propensity weight estimation method on uncertainty and the performance of analytic and bootstrap variance estimates. Standard error estimates for the four different propensity weight estimation methods are reported in Figure \ref{fig:sesim}.  For the analytic variance estimate, we used the design based errors when using weights estimated with CBPS, EB, and RF methods and the proposed analytic standard error estimate ($\widehat{V}_{Prop}$) when using weights estimated via logistic regression. The average analytic standard error estimate was generally comparable to the empirical standard error across simulations, but the bootstrap generally overestimated the true uncertainty.  The design based and proposed variance estimates resulted in similar standard error estimates.  Compared to the bootstrap estimate, analytic estimates more closely approximated the empirical standard error and were computationally easier than the bootstrap. 

In 21 out of 1,000 simulations, coefficients were not estimable in some bootstrap samples due to extreme values of estimated weights. There was one bootstrap sample in two simulations where the association was inestimable when weights were estimated using EB. This was slightly more common when weights were estimated using RF--out of 1,000 simulations, 19 simulations had at most 5 bootstrap sample that were unable to estimate the association.  We hypothesize that this occurs due to uniquely sparse bootstrap samples with little to no representation of some subpopulations, but this does not occur for logistic regression or CBPS. The RF method draws bootstrap samples to fit each tree and this bootstrap within a bootstrap may lead to subpopulations without any variation in the response, and thus extreme weights. Entropy balancing targets covariate balance which can be difficult if certain subpopulations are only observed in either the representative or convenience sample. Although CBPS also targets covariate balance, we did not observe any extreme weights and we hypothesize the focus on maximizing the likelihood may prevent this. Although this occurs infrequently, it may arise in practice. We excluded any bootstrap samples with insufficient information for estimating the association for a given sampling weight estimation method from the bootstrapped estimate of the variance.

\section{Application to the C2C Willingness Analysis}
\label{s:application}

The work of \cite{salazar_Racial_2020} investigated differences in research willingness by race and ethnicity using logistic regression. 
The methods are described in detail in that paper and we summarize them here. 
They performed logistic regression models to assess racial/ethnic differences in willingness to participate in research.  They separately evaluated 9 different responses: willingness to be contacted about studies involving (1) physical activity/diet modification, (2) cognitive testing, (3) magnetic resonance imaging (MRI), (4) positron emission tomography (PET) scans, (5) blood draws, (6) approved medications, (7) investigational medications, (8) lumbar punctures and (9) autopsy. They adjusted for age, sex, educational attainment, number of comorbidities, number of medications, cognitive function instrument score (\cite{amariglio_tracking_2015}, \cite{walsh_adcs_2006}) and research attitudes questionnaire score (\cite{rubright_measuring_2011}) and used multiple imputation to handle missing data. C2C data is updated as more participants enroll and can be requested at \url{https://c2c.uci.edu/request-c2c-data/}. When replicating this analysis, we started with the same dataset of C2C participants, but performed our own multiple imputation.   

\subsection{Identifying Matching Covariates}
\label{s:identifycov}
We first identified covariates likely to modify the relationship between race/ethnicity and willingness to participate in research and were collected in both the C2C and NHANES. Some covariates were recorded with differing levels of granularity in the two datasets so we collapsed them into comparable subgroups. 
For example, the question regarding exercise was phrased differently in the two datasets. NHANES participants were asked if they participated in vigorous or moderate recreational activities in a typical week for at least 10 minutes.The C2C participants were asked if they participated in the following activities for at least 15 minutes/day at least once/week for the last year: walking, hiking, biking, aerobics, calisthenics, swimming, water aerobics, weight training, stretching, or another form of exercise. We decided to exclude the question about walking for the C2C subjects because there were high agreement rates for this question and we were concerned participants may have reported walking for purposes other than recreational exercise. To compare across groups, we created an indicator for exercise. 
In total, we included 14 variables: age, sex, education level [Educ.] (less than 12 years [$<12$], high school/GED [12], some college [12-16], college graduate [16]), 
race/ethnicity (NH White, Hispanic, NH Asian, NH Black), high blood pressure (BP), kidney disease, liver disease, congestive heart failure (CHD), past cancer diagnosis, major depression, average hours of sleep per night, prescription medicine use (Presc. meds), and exercise.

\begin{figure}
    \centering
    \includegraphics[width=\textwidth]{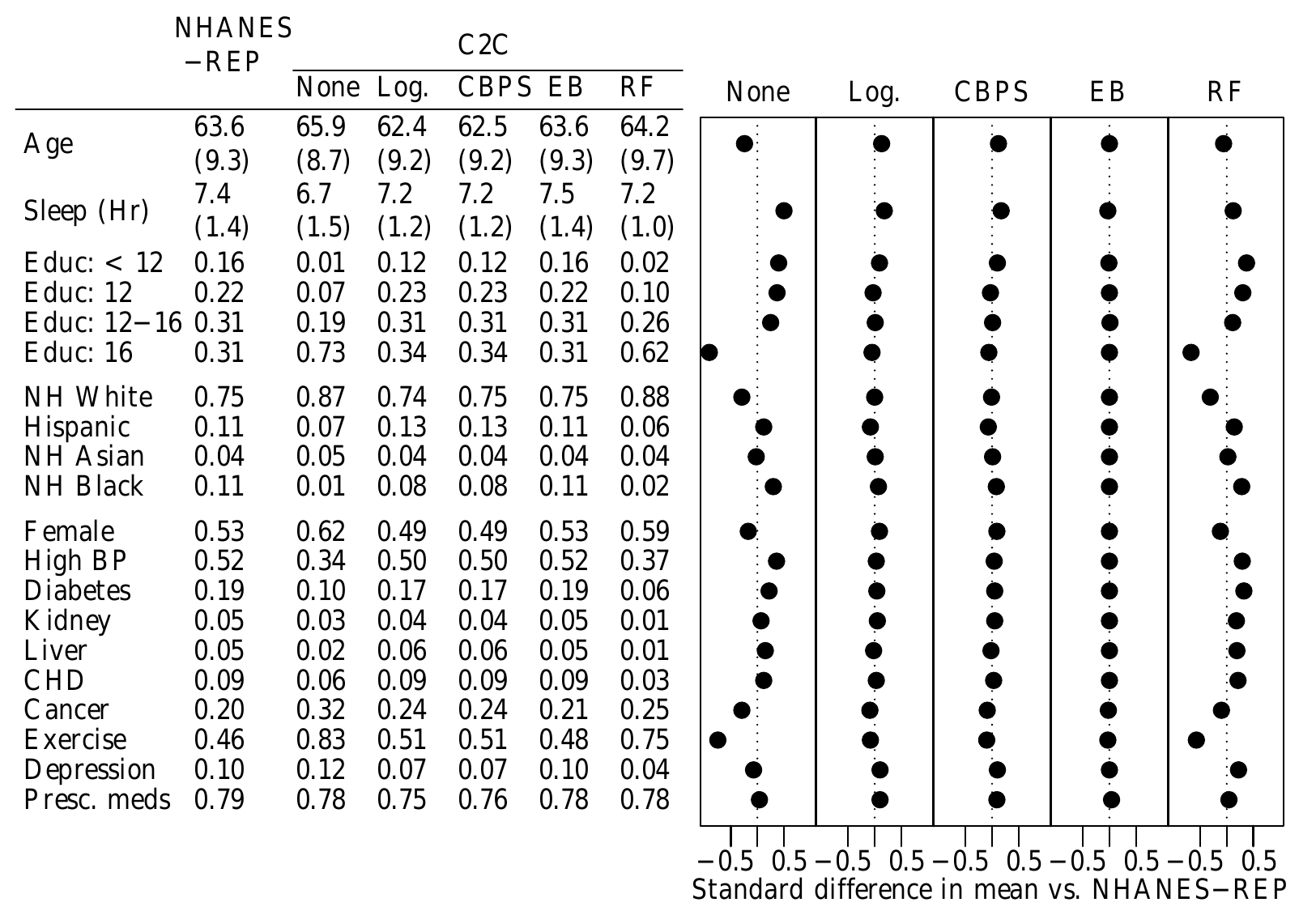}
    \caption{Covariates were summarized as mean (standard deviation) for continuous variables and as proportions for categorical variables in the table on the left for NHANES-REP (38,811 observations) and the C2C (2,749 subjects) datasets as well as a weighted C2C dataset. Weights were estimated using a logistic (Log.), covariate balancing propensity score (CBPS), entropy balancing (EB), or random forest (RF) method. Continuous covariates were summarized by mean (standard deviation) and categorical covariates by proportion. Standardized difference in means relative to NHANES-REP are presented in the figure on the right (\cite{stuart_matching_2010}). Propensity weights were estimated with missing values in the C2C imputed once. Covariates are described in Section \ref{s:identifycov}.}
\label{fig:covariates}
\end{figure}

Excluding subjects with a reported race or ethnicity of ``other" resulted in $n_C = 2,749$ observations in the C2C and excluding 917 NHANES participants with missing data (out of $n_S = 5,605$) resulted in $n_R = 38,811$ observations in NHANES-REP. Most of the missingness in the C2C was confined to a few covariates: 576 subjects were missing sleep, 207 were missing prescription drugs, 167 were missing exercise, and 29 were missing history of cancer.  Matching covariates for each dataset are summarized in Figure \ref{fig:covariates}. Continuous covariates were summarized by mean (standard deviation) and the proportion was reported for categorical variables. Weighted sample statistics using estimated propensity weights for the C2C dataset were also presented. The propensity weights were estimated using logistic regression, CBPS, EB, and RF with one imputed C2C dataset. Logistic regression, CBPS, and EB balance the covariate distributions well, but RF weighted estimates are similar to unweighted ones.

\subsection{Estimating propensity weights}

We repeated the analysis performed by Salazar et al.and imputed 5 C2C datasets and used Rubin's rules to aggregate across datasets (\cite{rubin_multiple_1987}). Within each dataset we estimated propensity weights for each subject using logistic regression, CBPS, EB, and RF. We fit the outcome models that quantify the relationship between race/ethnicity and each of the 9 outcomes and incorporated the estimated propensity weights. We report the estimated odds ratios (OR) and 95\% confidence intervals using the analytic variance estimates for each racial/ethnic group for each of the 9 responses. We report the full results with no weighting and each of the four types of propensity weights in Appendix \ref{s:fulltable} (Table \ref{tab:results}). A selection of these results are depicted using forest plots to discuss the impact of weighting.

\begin{figure}
    \centering
    \includegraphics[width=\textwidth]{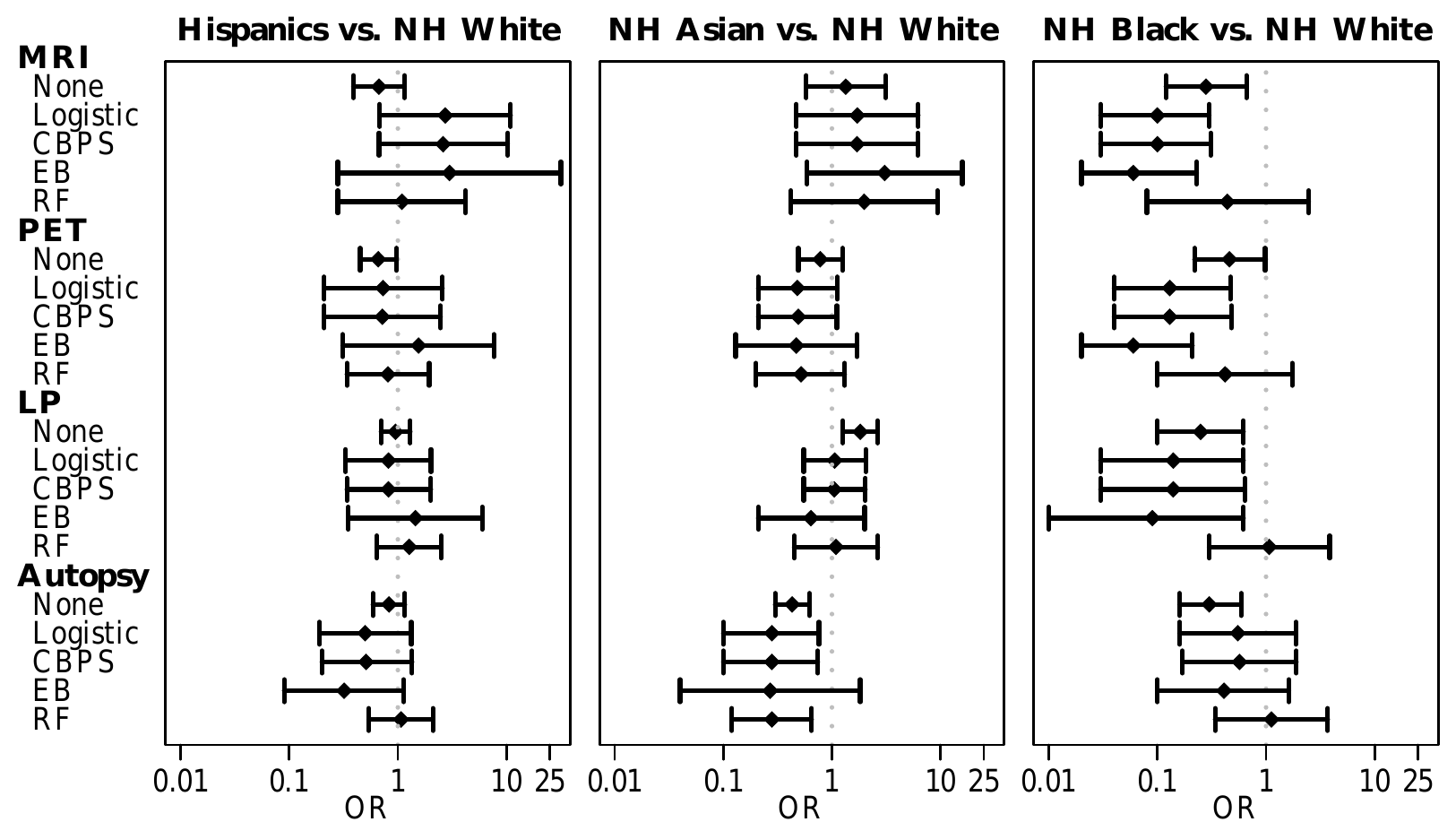}
    \caption{{Forest plots of the estimated odds ratios (OR) and 95\% confidence intervals for the racial ethnic differences analysis with MRI, PET, lumbar puncture (LP) and autopsy as the response. Results are presented for unweighted analysis (None) along with the weighted analysis using propensity weights estimated with logistic, covariate balancing propensity score (CBPS), entropy balancing (EB), and random forest (RF) methods.}}
    \label{fig:forest}
\end{figure}

Across all forest plots (Figure \ref{fig:forest}),
we observed that the standard errors increased with weighting, but this added variance better reflects the true uncertainty in the estimates and their ability to generalize to an external population. For example, the odds ratio comparing Hispanics to NH Whites for willingness to be contacted about studies with lumbar puncture (LP) had a noticeably wider confidence interval for the weighted estimates. The C2C underrepresents Hispanic subjects relative to the US population and the wider confidence intervals reflect this lack of information on the subpopulation. Several statistically significant odds ratios were no longer significant after incorporating estimated propensity weights. NH Asians had significantly higher odds of being willing to be contacted about studies involving LP compared to NH Whites in the original analysis, but this relationship was no longer statistically significant after weighting. \cite{salazar_Racial_2020} found it surprising that NH Asians would be more willing to undergo an LP because previous studies had found them less willing relative to NH Whites (\cite{moulder_factors_2017}). They speculated that NH Asians in the C2C had been exposed to more education about the LP procedure through outreach events for older Chinese adults. Accounting for sampling bias with estimated weights has attenuated this relationship to the null which aligns with previous findings.

The models using logistic and CBPS estimated weights tended to have similar estimates and confidence intervals. The CBPS model uses a logistic regression model but incorporates moment balancing conditions into the model fitting. These additional constraints did not impact the final result substantially when compared to the standard logistic regression derived estimates. The estimates using RF and EB weights had high variability and differed from the results using logistic and CBPS weights. Additionally, the point estimates from the models using RF weights tended to differ the most from the other 3 weighted models. Random forest is unique because it is both non-parametric and does not target covariate balancing. An advantage of decision trees is they naturally include interactions in modeling, but in sparse data with little representation of subpopulations this can lead to increased variability. Weight trimming, where estimated probabilities of 0 or 1 were replaced with 0.01 and 0.99, may have also impacted bias and variance estimates as well as the population of inference (\cite{lee_weight_2011}).


Incorporating estimated propensity weights not only impacted the uncertainty, it also changed the direction of point estimates. For example, in the original analysis Hispanics had lower odds of being willing to be contacted about studies that involve an MRI scan, but most of the weighted point estimates suggested Hispanics may actually have higher odds. The results for the original analysis were close to being statistically significant but the weighted models showed little evidence of an effect which leads to a different interpretation of the results. The four weighted point estimates were not covered by the unweighted confidence interval. 

\section{Discussion}
\label{s:discuss}

In our empirical results we demonstrated that including propensity weights estimated using NHANES can lead to better estimates of parameters pertaining to a general target population. Convenience samples are widely available and often used in research studies, but failing to account for the selection mechanism can lead to biased estimates and underestimation of the true variance of estimates. It is important to carefully select a target population and design studies and analyses that generalize to this population. If researchers are not able to obtain a representative sample because of ethical or practical considerations, they are forced to use a convenience sample. Since estimated propensity weights can only balance a convenience sample on observed covariates, researchers must take care to collect any covariates that they hypothesize are related to both the outcome and sampling probability. Additionally, any subpopulation with a convenience sample membership probability of zero and thus not represented in the convenience sample cannot be included in the target population. Researchers must carefully consider which covariates should be collected and which subpopulations are being sampled into a convenience sample to allow for valid estimates of associations in the desired target population. 

In the analysis of racial/ethnic differences of research willingness, weighted confidence intervals were generally at least twice as wide as unweighted confidence intervals. Incorporating propensity weights can increase the variability of parameter estimates because subjects with a low estimated probability of convenience sample membership will have large estimated weights and undue influence on the estimated associations (\cite{little_Complete-case_2014}). 
Although propensity weights may increase variance, they can reduce bias of the estimated association in the population of interest. Using an unrepresentative sample provides less information about the target population, and thus the increased variance of our estimator reflects this uncertainty (\cite{lumley_Complex_2010}). To measure how much the sampling mechanism impacts efficiency, we can compute the design effect, which is the ratio of the variance of the parameter estimate in an unrepresentative sample compared to a simple random sample (\cite{lumley_Complex_2010}, \cite{kish_survey_1965}). When comparing empirical variances for logistic-weighted coefficient estimates fit in a biased sample compared to unweighted estimates in a SRS, the design effect ranges from 1 to 1.6. Thus, we will need a biased sample that is up to 1.6 times bigger than a simple random sample to obtain the same variance. 

For estimating the variance of parameter estimates in the outcome model, we compared a resampling approach and a novel analytic approach for a logistic regression model that accounts for uncertainty arising from the propensity weight estimation. In empirical studies, we found the proposed analytic estimates performed better than the bootstrap estimate. Surprisingly, the bootstrap estimate tends to overestimate the uncertainty even though we replicated the estimation method within each bootstrap sample. Perhaps there was more variability in estimated weights within bootstrap samples than within the full simulated data set. Previous work on implementing entropy balancing weights (\cite{vegetabile_nonparametric_2021}) and exact matching using the propensity score (\cite{austin_estimating_2015}) also reported conservative bootstrap variance estimates, but the stratified double bootstrap where units are resampled from the survey sample and convenience sample used by \cite{ackerman_generalizing_2021} provides similar variance estimates to a design-based approach. The proposed analytic variance for model-based propensity weight estimation methods accounts for uncertainty in the estimated weights. The design based standard errors, however, perform well even though they fail to account for the propensity weight estimation process. We suggest using the proposed variance estimator with a model based propensity weight estimation procedure because it may perform better than the design based estimate, but the design based approach should perform well if needed. In our context, the two methods did not diverge substantially, but they could if there is high variability in the propensity weights. 
It is possible to derive an analytic variance estimator that accounts for the weight estimation for the CBPS model. However, it is made difficult because the CBPS model is overspecified and fit using generalized method of moments (\cite{imai_Covariate_2014}, \cite{hansen_large_1982}). One might be able to incorporate the final scientific model into the CBPS model as an additional balancing constraint for a simultaneous estimation approach. This may be an interesting area of research to pursue. 

All four propensity weight estimation models decreased bias in the simulation study. Algorithmic propensity weight estimation methods are very flexible, but random forest provided the smallest degree of bias reduction and the largest variance in the simulation study. Using RF-derived weights provided the poorest covariate balance (Figure \ref{fig:covariates}). Models that incorporate covariate balancing into model-selection help ensure covariate balance in the biased sample. EB, however, scales better than forward step-wise model selection for logistic regression, but CBPS tends to be slower due to the additional constraints. Estimates using CBPS did not deviate substantially from those using logistic regression, so the additional balancing constraints did not improve performance. We used the default settings for the \texttt{CBPS} package and users can change the settings to focus more on covariate balance. Although EB balanced covariates better than the logistic model in the applied example, they both reduced bias in estimated associations to the same extent in simulation study. Likelihood-based regression models, such as logistic regression, allow for an analytic variance estimate that fully accounts for uncertainty. It can also be easily expanded to include interaction and smoothers to allow for greater flexibility, but the second order terms seemed to perform well enough in our experiments. In practice, we suggest using logistic regression because it effectively reduces bias under our assumptions, is familiar to many scientists, is broadly accessible in different statistical software packages, and allows for an analytic variance estimate that accounts for uncertainty from estimation of propensity weights.



NHANES is a practical choice for generalizing biomedical studies to the US population. Different research areas may, however, collect variables that are not recorded in NHANES but are believed to be strongly related to the sampling probability. Other national surveys collect different variables and may be more relevant to different research areas.  Other examples of national surveys are the American Community Housing Survey which collects population and housing information, the Behavioral Risk Factor Surveillance System which conducts health-related telephone interviews, the General Social Survey which studies American society, and the Current Population Survey that collects labor force statistics. Additionally, researchers may want to generalize to a population outside of the US. If the target population is a subset of the US population, NHANES can be subset and used as the representative sample. Otherwise, other representative samples  need to be obtained. Researchers can consider census data if available, government sponsored national surveys, or international surveys. After specifying the target population, one should consider which samples are most representative and accessible.


Overall, estimated propensity weights reduce bias on parameter estimates from unrepresentative sampling. We, of course, are unable to account for any unmeasured covariates that may contribute to selection bias. Additionally, we collapsed different variables to match across different datasets and were unable to empirically evaluate if these are equivalent definitions.
Implementing estimated propensity weights increases the uncertainty of estimates but this reflects the information available on target population parameters. 

Convenience samples are easily collected and are used for research in many disciplines. 
The NHANES dataset is a rich, open access dataset that will likely have many overlapping covariates with convenience samples. Estimated propensity weights using NHANES is practical and effective at addressing selection bias concerns in convenience samples when trying to generalize to the non-institutionalized US population.

\appendix
\section{Derivation of Variance Estimator}
\label{s:mathderiv}

In this section we derive the various components of $I$ and $Q$ introduced in Section \ref{s:methodsvar}. Using iterated expectations we can show the cross-term $R$ is
\begin{align*}
R = E \Big[  \sum_{i\in \mathcal{C}}\bar{U}_i \Big(\sum_{i\in \mathcal{C}}T^T_i + \sum_{i\in \mathcal{R}}T^T_i)\Big) \Big|X=x,Z=z\Big] = E\Big[  \sum_{i\in \mathcal{C}}\bar{U}_i \sum_{i\in \mathcal{C}}T^T_i \Big|X=x,Z=z\Big].
\end{align*}
So the method-of-moments estimator for $Q$ is
\begin{align*}
    \widehat{Q} = \begin{pmatrix}
    TT^T & \widehat{R}^T \\
    \widehat{R} &
    \bar{U}\bar{U}^T 
    \end{pmatrix}\Bigg|_{\binom{\beta}{\gamma} = 
    \binom{\widehat\beta}{\widehat\gamma}}
\end{align*}
with $\widehat{R} = \sum_{i\in \mathcal{C}}\bar{U}_i \sum_{i\in \mathcal{C}}T_i^T$.\\

Now consider the the terms of $I$. $I_{TT}$ is the derivative of T with respect to $\gamma$ which is Fisher's information matrix for logistic regression,
\begin{align*}
    I_{TT} = \sum_{i\in\mathcal{C}\cup\mathcal{R}} E \Big(-\frac{\partial T_i}{\partial \gamma}\Big| X_i = x_i\Big) 
    =\sum_{i\in\mathcal{C}\cup\mathcal{R}} \Big( {x}_{i}^T (P_{Ci}(1-P_{Ci})) {x}_{i} \Big).
\end{align*}
$I_{UU}$ is similar,
\begin{align*}
    I_{UU} = -\sum_{i\in\mathcal{C}} E\Big[ w_{Ci}\frac{\partial U_j}{\partial \beta_k} \Big| Z = z\Big ] 
    = z^T M(\beta) z 
\end{align*}
where
\begin{align*}
    M(\beta) = diag \Big(\frac{w_{Ci} (\partial \eta_i/ \partial\mu_i)^{-2}}{V(\mu_i|Z_i=z_i)} \Big ).
\end{align*}
Finally, $I_{UT}$ is
\begin{align*}
    I_{UTkj} = - \sum_{i\in\mathcal{C}\cup\mathcal{R}}E\Big[ \frac{\partial U_{ki}}{\partial \gamma_{j}} \Big| X_i = x_i, Z_i=z_i\Big ] 
    = \sum_{i\in\mathcal{C}\cup\mathcal{R}}E\Big[w_{Ci}\Big (\frac{\partial \mu_i}{\partial \beta_j} \Big) \Big(\frac{Y_i - \mu_i}{V(\mu_i)} \Big)X_{ij} \Big| X_i = x_i, Z_i=z_i\Big] .
\end{align*}

We can rearrange $I^{-1}Q I^{-1}$ to point out the relationship to the design based variance used in the survey sampling literature (\cite{lumley_fitting_2017}). Define $\widehat{A} = \widehat{I}_{UU}$ and $\widehat{B} = \bar U \bar U^T$ so that,
\begin{align*}
    \widehat{Q} = \begin{pmatrix}
    T \\ \bar{U}
    \end{pmatrix}
    \begin{pmatrix}
    T^T \bar{U}^T
    \end{pmatrix} 
    = \begin{pmatrix}
    TT^T & \widehat{R}^T \\
    \widehat{R} & \widehat{B}
    \end{pmatrix} .
\end{align*}
Using the formula for blockwise inversion (Fact 2.17.1 in \cite{bernstein_basic_2009}),
\begin{align*}
\setlength\arraycolsep{4pt}
    I^{-1} = 
    \begin{pmatrix}
    I_{TT} & 0 \\
    I_{UT} & A
    \end{pmatrix} ^{-1}
    = 
    \begin{pmatrix}
    I_{TT}^{-1} && 0 \\
    -A^{-1}I_{UT}I_{TT}^{-1} && A^{-1}
    \end{pmatrix}.
\end{align*}
Combining $I^{-1}$ and $Q$ we obtain the proposed variance estimator, 
\begin{align*}
   \widehat{V}_{Prop}(\widehat\beta) &=  \widehat{A}^{-1}\widehat{B}\widehat{A}^{-1} - \widehat{A}^{-1}\widehat{I}_{UT}\widehat{I}_{TT}^{-1}\widehat{R}^T \widehat{A}^{-1}.
\end{align*}

\pagebreak

\section{Bias Adjusted C2C Results}
\label{s:fulltable}
\begin{table}[!h]
\centering
\caption{Bias adjusted C2C results: Odds Ratios and 95\% confidence intervals are presented for the models from \cite{salazar_Racial_2020} assessing the relationship between race/ethnicity and 9 responses with adjustment variables. The models were fit without any propensity weights and with propensity weights estimated using logistic regression, covariate balancing propensity score (CBPS), entropy balancing (EB), and random forest (RF) methods.}
\label{tab:results}
\resizebox{.75\textwidth}{!}{
\begin{tabular}{lclll}
\toprule
Trial type & Model & Hispanic & NH Asian & NH Black\\
\midrule
Physical Activity / & Unweighted & 1.06 (0.52, 2.16) & 0.68 (0.34, 1.35) & 1.87 (0.25, 13.95)\\
Diet Modification & Logistic & 1.52 (0.40, 5.83) & 0.82 (0.23, 2.97) & 4.54 (0.45, 45.38)\\
 & CBPS & 1.52 (0.40, 5.76) & 0.83 (0.23, 3.02) & 4.64 (0.46, 46.28)\\
 & EB & 4.80 (0.54, 42.33) & 0.35 (0.06, 2.18) & 4.61 (0.31, 69.07)\\
 & RF & 0.74 (0.16, 3.50) & 0.87 (0.19, 4.09) & 3.93 (0.36, 43.02)\\
\addlinespace
Cognitive Testing & Unweighted & 0.50 (0.22, 1.12) & 0.52 (0.18, 1.52) & 0.71 (0.09, 5.55)\\
 & Logistic & 0.73 (0.13, 4.26) & 0.47 (0.06, 3.89) & 0.95 (0.09, 10.14)\\
 & CBPS & 0.72 (0.13, 4.04) & 0.45 (0.05, 3.72) & 0.99 (0.09, 10.66)\\
 & EB & 1.75 (0.09, 33.19) & 0.91 (0.03, 23.98) & 0.93 (0.08, 11.60)\\
 & RF & 0.27 (0.06, 1.27) & 0.88 (0.15, 5.36) & 1.41 (0.13, 15.37)\\
\addlinespace
MRI Scans & Unweighted & 0.67 (0.39, 1.15) & 1.34 (0.58, 3.12) & 0.28 (0.12, 0.66)\\
 & Logistic & 2.73 (0.68, 10.90) & 1.71 (0.47, 6.22) & 0.10 (0.03, 0.30)\\
 & CBPS & 2.61 (0.67, 10.25) & 1.70 (0.47, 6.17) & 0.10 (0.03, 0.31)\\
 & EB & 2.99 (0.28, 31.66) & 3.06 (0.59, 15.86) & 0.06 (0.02, 0.23)\\
 & RF & 1.09 (0.28, 4.19) & 1.98 (0.42, 9.35) & 0.44 (0.08, 2.47)\\
\addlinespace
PET Scans & Unweighted & 0.66 (0.45, 0.97) & 0.78 (0.49, 1.25) & 0.46 (0.22, 0.98)\\
 & Logistic & 0.73 (0.21, 2.57) & 0.48 (0.21, 1.12) & 0.13 (0.04, 0.47)\\
 & CBPS & 0.72 (0.21, 2.47) & 0.49 (0.21, 1.11) & 0.13 (0.04, 0.48)\\
 & EB & 1.55 (0.31, 7.65) & 0.47 (0.13, 1.70) & 0.06 (0.02, 0.21)\\
 & RF & 0.81 (0.34, 1.94) & 0.52 (0.20, 1.31) & 0.42 (0.10, 1.74)\\
\addlinespace
Blood Draws & Unweighted & 0.62 (0.35, 1.10) & 0.31 (0.18, 0.53) & 0.27 (0.11, 0.67)\\
 & Logistic & 1.62 (0.45, 5.75) & 0.37 (0.15, 0.89) & 0.70 (0.15, 3.20)\\
 & CBPS & 1.58 (0.45, 5.54) & 0.37 (0.15, 0.88) & 0.70 (0.16, 3.16)\\
 & EB & 4.17 (0.69, 25.15) & 0.24 (0.05, 1.07) & 0.61 (0.12, 3.01)\\
 & RF & 0.77 (0.17, 3.47) & 0.16 (0.04, 0.60) & 0.29 (0.05, 1.61)\\
\addlinespace
Approved & Unweighted & 0.68 (0.42, 1.10) & 0.61 (0.36, 1.01) & 0.67 (0.25, 1.80)\\
Medications & Logistic & 0.17 (0.06, 0.49) & 0.66 (0.29, 1.50) & 0.72 (0.21, 2.40)\\
 & CBPS & 0.17 (0.06, 0.49) & 0.65 (0.29, 1.49) & 0.72 (0.21, 2.41)\\
 & EB & 0.13 (0.02, 0.79) & 0.43 (0.08, 2.44) & 0.81 (0.10, 6.69)\\
 & RF & 0.64 (0.19, 2.17) & 0.44 (0.15, 1.29) & 0.45 (0.14, 1.45)\\
\addlinespace
Investigational & Unweighted & 0.62 (0.42, 0.90) & 0.55 (0.36, 0.83) & 0.52 (0.24, 1.11)\\
Medications & Logistic & 0.31 (0.10, 0.93) & 0.33 (0.14, 0.77) & 0.70 (0.25, 1.98)\\
 & CBPS & 0.31 (0.10, 0.93) & 0.33 (0.14, 0.76) & 0.71 (0.25, 1.98)\\
 & EB & 0.50 (0.07, 3.73) & 0.29 (0.08, 1.04) & 0.82 (0.25, 2.75)\\
 & RF & 0.67 (0.26, 1.67) & 0.40 (0.15, 1.08) & 0.36 (0.11, 1.15)\\
\addlinespace
Lumbar Puncture & Unweighted & 0.95 (0.70, 1.30) & 1.82 (1.26, 2.63) & 0.25 (0.10, 0.62)\\
 & Logistic & 0.82 (0.33, 2.02) & 1.06 (0.55, 2.05) & 0.14 (0.03, 0.62)\\
 & CBPS & 0.82 (0.34, 2.01) & 1.05 (0.55, 2.03) & 0.14 (0.03, 0.64)\\
 & EB & 1.45 (0.35, 5.97) & 0.64 (0.21, 2.00) & 0.09 (0.01, 0.62)\\
 & RF & 1.27 (0.64, 2.50) & 1.09 (0.45, 2.64) & 1.07 (0.30, 3.85)\\
\addlinespace
Autopsy & Unweighted & 0.83 (0.59, 1.16) & 0.43 (0.30, 0.62) & 0.30 (0.16, 0.59)\\
 & Logistic & 0.50 (0.19, 1.33) & 0.28 (0.10, 0.76) & 0.55 (0.16, 1.88)\\
 & CBPS & 0.51 (0.20, 1.35) & 0.28 (0.10, 0.74) & 0.57 (0.17, 1.90)\\
 & EB & 0.32 (0.09, 1.13) & 0.27 (0.04, 1.82) & 0.41 (0.10, 1.62)\\
 & RF & 1.07 (0.54, 2.10) & 0.28 (0.12, 0.65) & 1.12 (0.34, 3.70)\\
\bottomrule
\end{tabular}}
\end{table}

\pagebreak

\section*{Acknowledgements}
The C2C Registry was made possible through a philanthropic gift from HCP, Inc and grants P30 AG066519 and UL1 TR000153. OMB was supported by the National Science Foundation Graduate Research Fellowship under Grant No. DGE-183928 and the ARCS foundation. BGV was supported by the National Institute on Drug Abuse of the National Institutes of Health under award number R01DA045049. CRS was supported by the National Institute on Aging of the National Institute of Health under a diversity supplement to award AG059407 and an Alzheimer’s Association research fellowship AARFD-20-682432. JDG and DLG were supported by the National Institutes of Health under award P30AG066519. DLG was also supported by the National Institute of Health under award R01AG053555. The content is solely the responsibility of the authors and does not necessarily represent the official views of the National Institutes of Health. 

\pagebreak

\bibliographystyle{apalike} 
\bibliography{references}       

\end{document}